\documentclass[11pt,twoside,onecolumn]{article}
\usepackage[]{latexsym}
\usepackage{epsfig}
\usepackage{amsmath,amssymb}
\newcommand{\be}{\begin{eqnarray}}
\newcommand{\ee}{\end{eqnarray}}

\title{\bf Charged Black Hole Remnants at the LHC}
\author{G.L.~Alberghi$^{a}$\thanks{alberghi@bo.infn.it},
$\ $
L.~Bellagamba$^{a}$\thanks{lorenzo.bellagamba@bo.infn.it},
$\ $
X.~Calmet$^{b}$\thanks{x.calmet@sussex.ac.uk},
$\ $
R.~Casadio$^{a,c}$\thanks{roberto.casadio@bo.infn.it},
$\ $
and
O.~Micu$^{d}$\thanks{octavian.micu@spacescience.ro}
\\
\null
\\
$^a${\em Istituto Nazionale di Fisica Nucleare, Sezione di Bologna}
\\
{\em viale B.~Pichat~6/2, 40127 Bologna, Italy}
\\
\\
$^b${\em Physics and Astronomy, University of Sussex}
\\
{\em Falmer, Brighton, BN1 9QH, UK}
\\
\\
$^c${\em Dipartimento di Fisica e Astronomia, Universit\`a di Bologna}
\\
{\em via Irnerio~46, 40126 Bologna, Italy}
\\
\\
$^d${\em Institute of Space Science}
\\
{\em P.O.Box MG-23, Ro 077125 Bucharest-Magurele, Romania}
}
\begin{document}
\maketitle
\begin{abstract}
We investigate possible signatures of long-lived (or stable) charged black holes
at the Large Hadron Collider.
In particular, we find that black hole remnants are characterised by quite low speed.
Due to this fact, the charged remnants could, in some cases, be very clearly distinguished 
from the background events, exploiting $dE/dX$ measurements. 
We also compare the estimate energy released by such remnants with that of typical
Standard Model particles, using the Bethe-Bloch formula.
\end{abstract}
\setcounter{page}{1}
\section{Introduction}
\setcounter{equation}{0}
It is now well appreciated that the scale at which quantum gravity effects
become comparable in strength to the forces of the Standard Model (SM)
of particle physics could be well below the traditional $10^{19}\,$GeV,
and potentially in the TeV~range~\cite{arkani,RS,Calmet:2008tn}.
In fact, models with low scale quantum gravity allow for a fundamental scale
of gravity as low as the electroweak scale, say $M_{\rm G}\simeq 1\,$TeV,
and microscopic black holes (BHs) may therefore be produced at the LHC,
the Large Hadron Collider (see, e.g., Ref.~\cite{cavaglia} for recent reviews).
Till recently, only semiclassical BHs, which  decay via the
Hawking radiation~\cite{hawking}, had been considered.
These BHs, whose standard description is based on the canonical Planckian
distribution for the emitted particles, have a very short life-time
of the order of $10^{-26}\,$s~\cite{dimopoulos}.
The creation of semiclassical BHs in collisions of high energetic particles
is well understood~\cite{penrose,D'Eath:1992hb,Eardley:2002re,Hsu:2002bd,sampaio}.
Our understanding of this phenomenon thus goes way beyond the naive
hoop conjecture~\cite{hoop} used in the first papers on the topic.
Several simulation tools are also available to describe semiclassical BH
production and evaporation at colliders~\cite{harris,catfish,cha2,blackmax,charybdis2},
and the LHC has already been able to set some bounds on the Planck mass
searching for semiclassical BHs~\cite{CMS,park}.
\par
Recently, it has been pointed out that, besides semiclassical BHs,
which appear to be difficult to produce at colliders, as they might require energies 5 to 20
times larger than the Planck scale, quantum BHs, could be instead copiously
produced~\cite{Calmet:2008dg,Calmet:2011ta,Calmet:2012cn,Calmet:2012mf}.
These BHs are non-thermal objects with masses close to the Planck scale,
and might resemble strong gravitational rescattering events~\cite{Meade:2007sz}. 
In Ref.~\cite{Calmet:2008dg}, non-thermal quantum BHs were assumed
to decay into only a couple of particles.
However, depending on the details of quantum gravity, the smallest quantum BHs
might be stable and would not decay at all.
Existence of remnants, i.e.~the smallest stable BHs, have been considered previously
in the literature~\cite{Koch:2005ks,Hossenfelder:2005bd}.
And, most recently, the production of neutral and integer charged semiclassical
remnant BHs have been simulated in Ref.~\cite{Bellagamba:2012wz}.
\par
In this work, we continue the phenomenological analysis of these events,
and study in particular the particle speed distributions, in search for a possible
clear signature to differentiate remnant BHs from SM particles.
For this purpose, we will again employ CHARYBDIS2~\cite{charybdis2},
as it is the only available code which can generate remnants, 
and will show that remnant BHs could be produced with relativistic factors
much smaller than those of SM particles. 
For the charged remnant BHs, we will also estimate the typical energy loss inside
a detector using the Bethe-Bloch equation and the distributions in speed.
\par
The paper is organised as follows: in the next Section, we briefly review BH production
in the semiclassical regime and extrapolate possible behaviours in the quantum regime,
mostly arguing on the possible existence of BHs with electric charge
(see also Appendix~\ref{BWBH});
in Section~\ref{charged}, we summarise the main results from Ref.~\cite{Bellagamba:2012wz},
and complete our analysis for remnant BHs with the study of their distributions in speed.
We confront these distributions both with the analogous distributions
for SM particles produced in the same events from partial BH decay
and with the distributions in events with the $t\,\bar t$ production (generated
using Powheg~\cite{powheg}  and Pythia~\cite{pythia} for the parton shower
and the hadronization), which represents one of the main backgrounds 
for processes of interest here.
We also estimate the typical energy released in a detector by charged remnants;
we finally comment and argue about further developments in Section~\ref{conc}.
\section{Black hole production}
\label{production}
\setcounter{equation}{0}
In this work, we start from the possibility that remnant BHs could not just be the end-point
of the Hawking evaporation, but they might also be produced directly without going
through the usual evaporation process.
The physics of the latter is very similar to that described in Ref.~\cite{Calmet:2008dg},
with the important exception that they would be stable.
\par
It is now well understood that semiclassical BHs (i.e.~thermal objects which decay
via Hawking radiation to many particles) will not be produced at the LHC.
The reason is simply that the mass of a BH needs to be several times larger than the
Planck mass $M_{\rm G}$ to be in the semiclassical regime.
In particular, the mass of the lightest semiclassical BHs is expected to be between
5 and 20 times $M_{\rm G}$, depending on the model.
Thus even if the Planck mass was in the few TeV region, one would not be able
to produce many semiclassical BHs at a $14\,$TeV LHC.
Following~\cite{Calmet:2008tn}, we consider quantum BHs that are produced directly
from the colliding particles.
However, while it was assumed in Ref.~\cite{Calmet:2008tn} that these holes would
totally decay into a few particles, we consider here the possibility that these holes 
can at most emit a fraction of their energy, and then become stable remnants,
corresponding to the lightest quantum BHs.  
\par
In a proton-proton collider such as the LHC, BHs would be produced
by quarks, anti-quarks and gluons and would thus typically carry both a QED
and a SU(3)$_c$ charge, namely
\begin{itemize}
\item[a)] ${\bf 3} \times {\bf \overline 3}= {\bf 8} + {\bf 1}$
\item[b)] ${\bf 3} \times {\bf 3}= {\bf 6} + {\bf \overline 3}$
\item[c)] ${\bf 3} \times {\bf 8}= {\bf 3} + {\bf \overline 6}+ {\bf 15}$
\item[d)] ${\bf 8} \times {\bf 8}= {\bf 1}_S + {\bf 8}_S+ {\bf 8}_A+{\bf 10} + {\bf \overline{10}}_A+ {\bf 27}_S$
\end{itemize}
Most of the time, BHs will thus be created with a SU(3)$_c$ charge and
come in different representations of SU(3)$_c$, as well as QED charges.
It is also likely that their masses are quantized, as described in Ref.~\cite{Calmet:2012cn}.
Quantum BHs can therefore be classified according to representations of SU(3)$_c$.
Let us further note that in Refs.~\cite{Calmet:2008dg,Calmet:2011ta,Calmet:2012cn,Calmet:2012mf}
we did not expect that non-thermal quantum BHs would ``hadronize" before decaying
(since the QCD length scale is 200$^{-1}\,$MeV, whereas that of quantum gravity
in these scenario is at most 1000$^{-1}\,$GeV).
However, since the BHs we consider here do not decay completely, we expect that they
will hadronize, i.e.~absorbe a particle charged under $SU(3)_c$ after traveling over a
distance of some $200^{-1}\,$MeV and become an $SU(3)_c$ singlet.
Or they could loose colour charge by emitting a fraction of their
energy before becoming stable.
In any case, the hadronization process could still lead to remnants with a (fractional) QED
charge and their phenomenology could be different from the one envisaged in
Ref.~\cite{Bellagamba:2012wz}.
To summarise, quantum BHs can be neutral or have the following QED charges:
$\pm 4/3$, $\pm 1$, $\pm 2/3$, and $\pm 1/3$.
Moreover, if the BH is fast moving, it is likely to hadronize in the detector, whereas
if it is moving slowly, this is likely to happen before it reaches the detector.
\par
The production cross section of quantum BHs is extrapolated from the semiclassical
regime and assumed to be accurately described by the geometrical cross section formula.
The horizon radius, which depends on the number $d$ of extra-dimensions, is given by
\be
R_{\rm H}
=\frac{\ell_{\rm G}}{\sqrt{\pi}}\,
\left(\frac{M}{M_{\rm G}}\right)^{\frac{1}{d+1}}
\left(\frac{8\,\Gamma\left(\frac{d+3}{2}\right)}{d+2}
\right)^{\frac{1}{d+1}}
\ ,
\ee
where $\ell_{\rm G}=\hbar/M_{\rm G}$ is the fundamental gravitational
length associated with $M_{\rm G}$,
$M$ is the BH mass, and $\Gamma$ the Gamma function.
At the LHC, a BH could form in the collision of two partons, i.e.~the quarks, anti-quarks
and gluons of the colliding protons.
The total BH cross section, 
\be
\left.\frac{d\sigma}{d M}\right|_{pp\to BH+X}
=\frac{dL}{d M}\,\sigma_{\rm BH}(ab\to BH; \hat s=M^2)
\ ,
\ee
can be estimated from the geometrical hoop conjecture~\cite{hoop},
so that 
\be
\sigma_{\rm BH}(M)\approx \pi\,R_{\rm H}^2
\ ,
\ee
and
\be
\frac{dL}{d M}=\frac{2\,M}{s}\,\sum_{a,b}\int_{M^2/s}^1
\frac{dx_a}{x_a}\,f_a(x_a)\,f_b\left(\frac{M^2}{s\,x_a}\right)
\ ,
\ee
where $a$ and $b$ represent the partons which form the BH,
$\sqrt{\hat s}$ is their centre-mass energy, $f_i(x_i)$ are
parton distribution functions (PDF), and $\sqrt{s}$ is the LHC centre-mass
collision energy (up to $8\,$TeV presently, with a planned maximum of $14\,$TeV).
\par
As an example, in the following we shall consider $\sqrt{s}=14\,$TeV,
and $\sigma_{\rm BH}\simeq 39.9\,$fb for $M_{\rm G}=3.5\,$TeV in $D=6$
dimensions~\cite{Bellagamba:2012wz}.
In the first few months of the future LHC run, a luminosity $L\simeq 10\,$fb$^{-1}$
should be reached, and one can therefore expect a total of about $400$ BH events.
\section{Remnant BHs detection at supercolliders}
\label{charged}
\setcounter{equation}{0}
\begin{figure}[t]
\centerline{\begin{tabular}{ cc}
\epsfig{figure=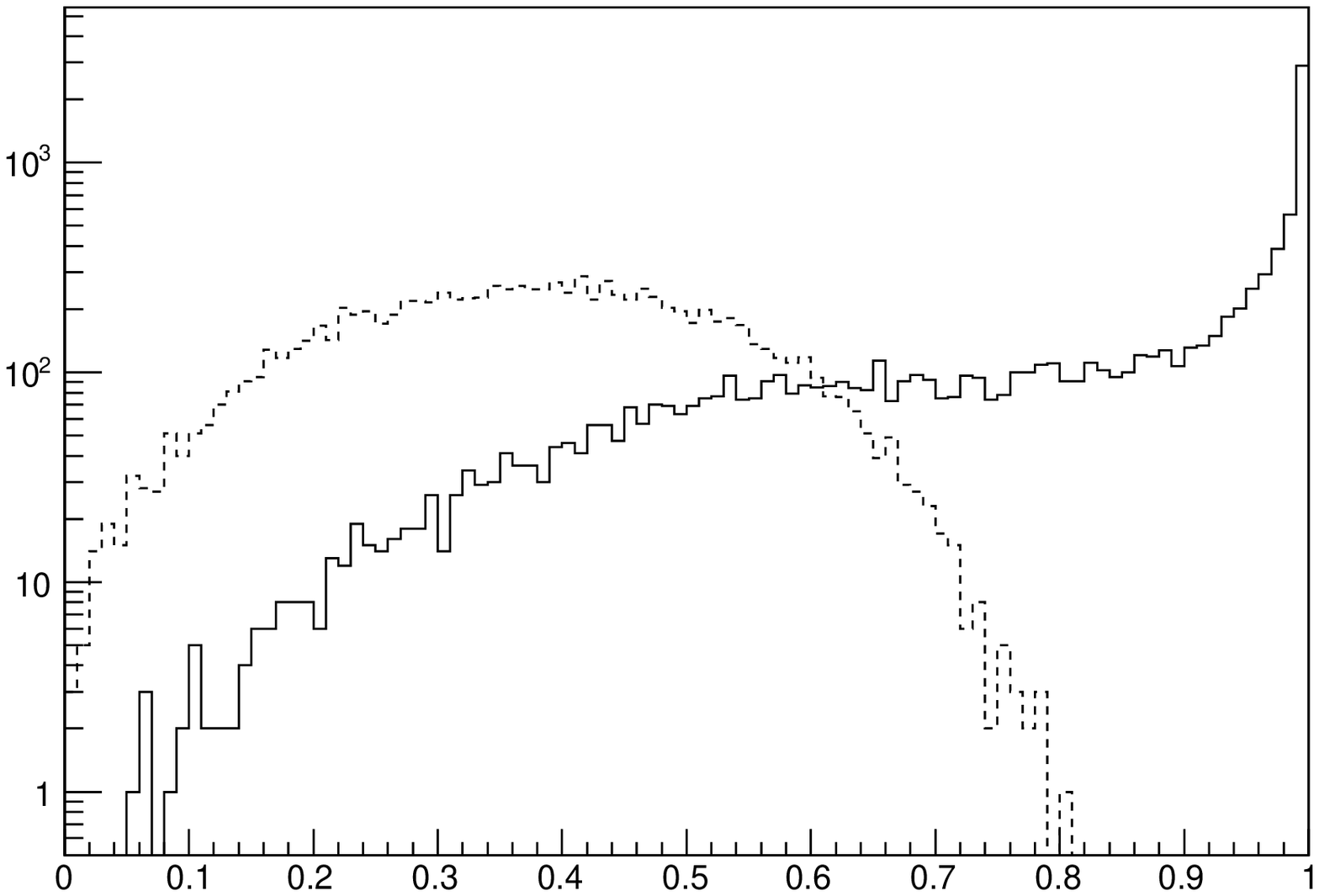,height=6cm,width=8cm}
&
\epsfig{figure=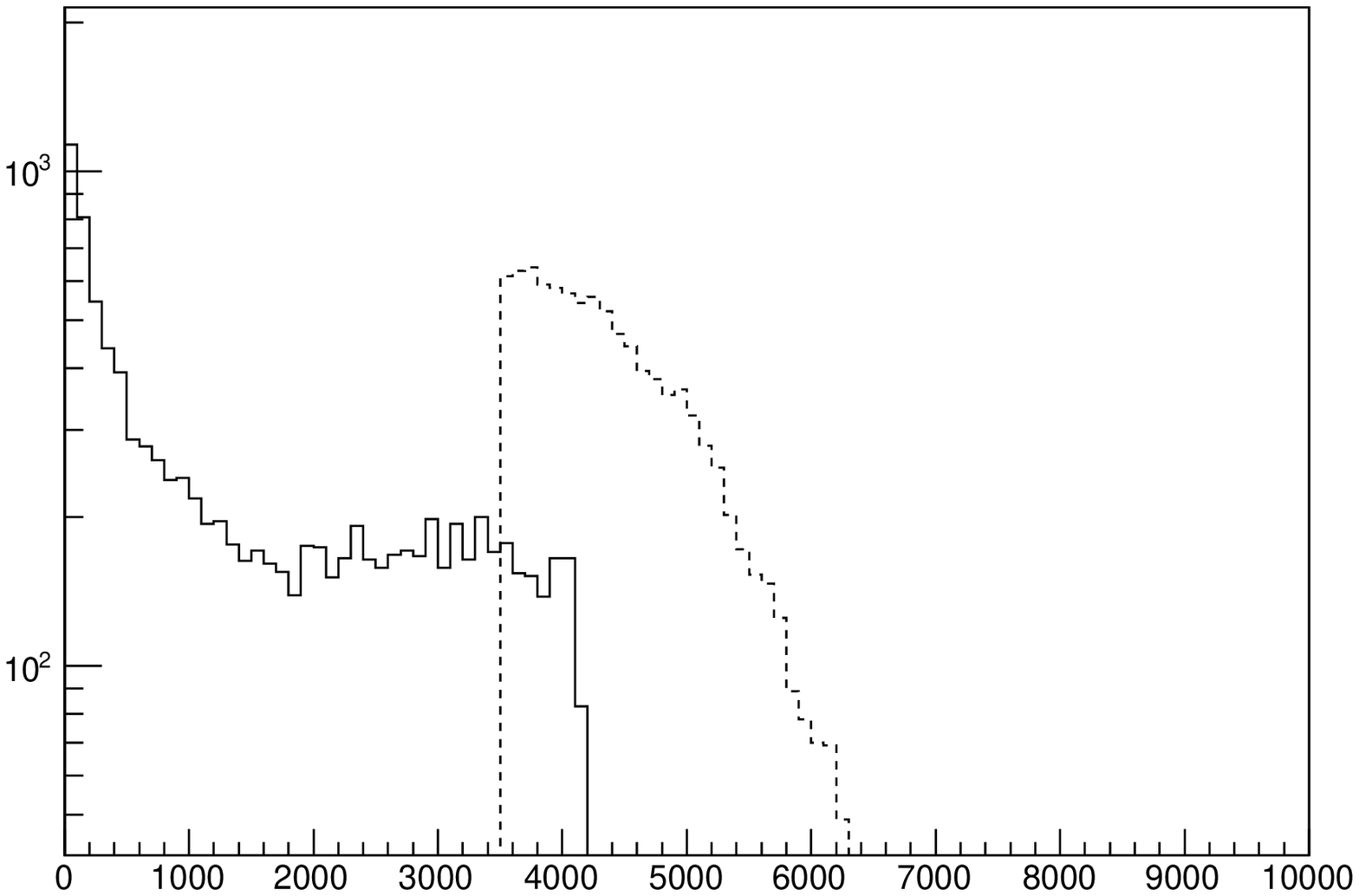,height=6cm,width=8cm}
\\
\small{$\beta_0$}
&
{\small{$M_0$}}
\end{tabular}
}
\caption{Distribution of speed $\beta_0$ (left panel) and mass $M_0$ (in GeV; right panel)
of the remnant BHs for KINCUT=TRUE (dashed line) and KINCUT=FALSE (solid line).
Both plots are for $\sqrt{s}=14\,$TeV with $M_{\rm G}=3.5\,$TeV and initial 
$M_{\rm BH}\ge 2\,M_{\rm G}$ in $D=6$ total dimensions and $10^4$ total BH events.
\label{beta}}
\end{figure}
\begin{figure}[h!]
\centerline{\begin{tabular}{ cc}
\epsfig{figure=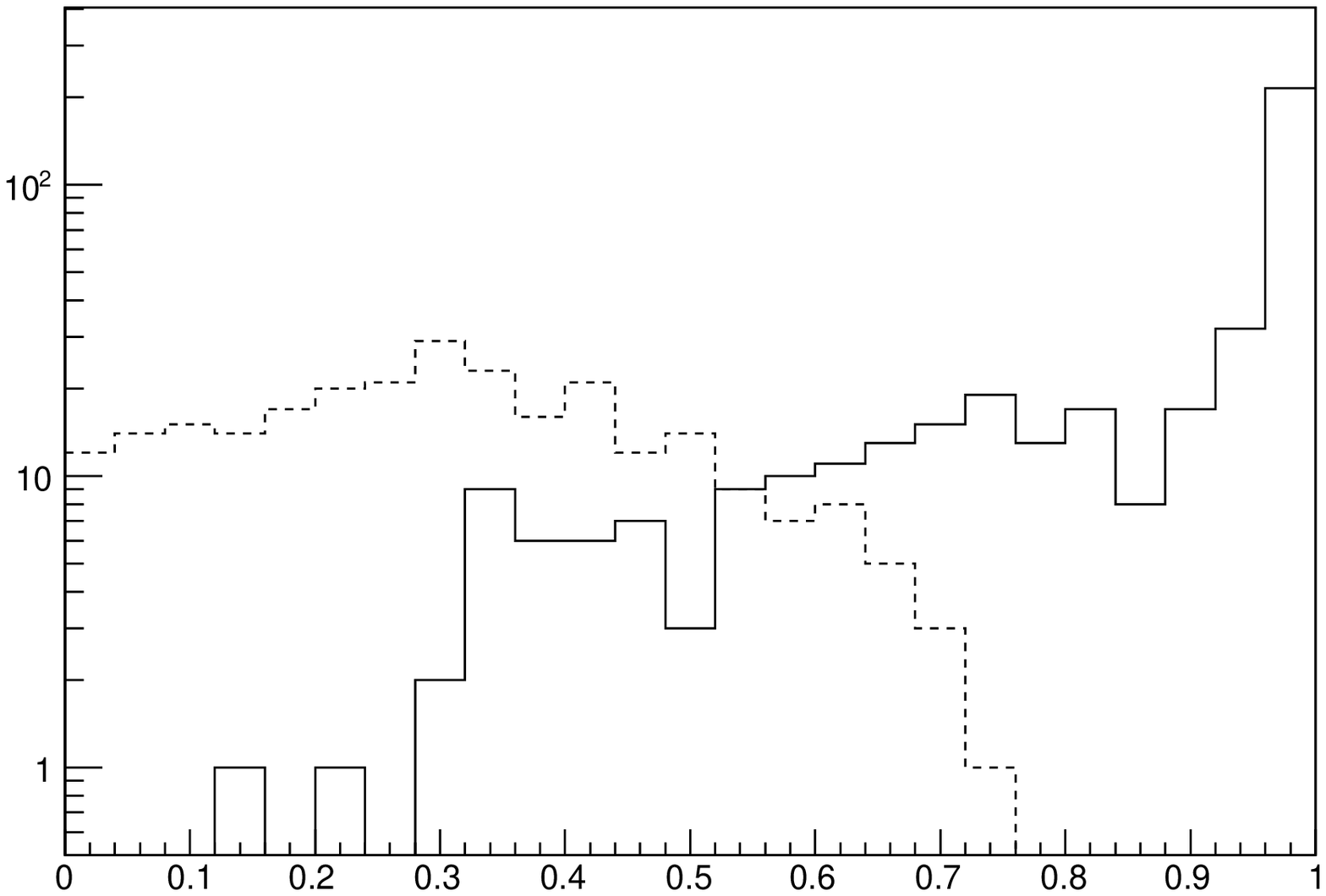,height=6cm,width=8cm}
&
\epsfig{figure=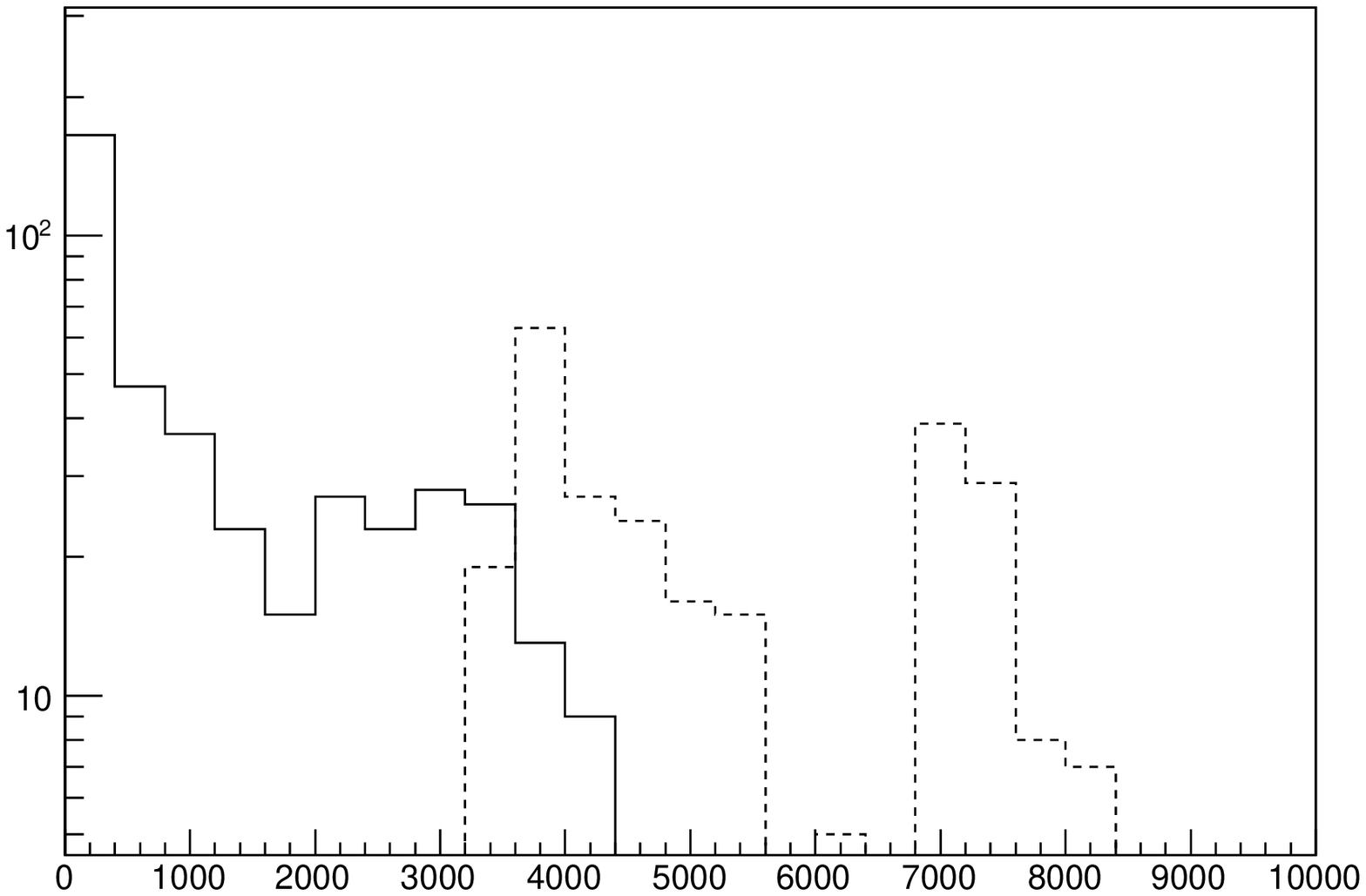,height=6cm,width=8cm}
\\
\small{$\beta_0$}
&
\small{$M_0$}
\end{tabular}}
\caption{Distribution of speed $\beta_0$ (left panel) and mass $M_0$ (in GeV; right panel)
of the charged remnant BHs for KINCUT=TRUE
(dashed line) and KINCUT=FALSE (solid line).
Both plots are for $\sqrt{s}=14\,$TeV with $M_{\rm G}=3.5\,$TeV in $D=6$
total dimensions and $10^4$ total events.
\label{betac}}
\end{figure}
The existence of semiclassical remnant BHs have been the subject of  
Monte Carlo simulations~\cite{Bellagamba:2012wz} employing the code
CHARYBDIS2~\cite{charybdis2}.
Such simulations have shown that a small percentage (of the order of $10\%$)
of the remnants will carry an electric charge $Q=\pm e$~\cite{Bellagamba:2012wz}.
\par
At present, there is no code specifically designed to simulate the phenomenology
of quantum BHs.
We thus employed CHARYBDIS2 to study quantum BHs, since they are produced
according to the same geometrical cross section as semiclassical BHs, and the details
of their possible partial decay are not phenomenologically relevant in searching for
a main signature of the existence of remnants.
In fact, for obvious kinematical reasons, the initial BH mass cannot be much larger
than a few times $M_{\rm G}$ (we typically set  $M_{\rm G}=3.5\,$TeV in our simulations),
even for $\sqrt{s}=14\,$TeV.
CHARYBDIS2 will hence make these BHs emit at most a fraction of their energy in a
small number of SM particles before they become stable.
Such a discrete emission process in a relatively narrow range of masses
is fundamentally constrained by the conservation of energy and the SM charges,
and cannot differ significantly for different couplings of the quantum BH to SM particles.
One limitation which occurs generically with Monte Carlo generators is that the decay
time is assumed to be instantaneous.
Our analysis does therefore not include the possibility that the BHs partially decay
off the production vertex, nor the effects of BH hadronization by absorption of coloured
particles which we theoretically discussed in Section~\ref{production}.
As in the case of semiclassical BHs~\cite{Bellagamba:2012wz}, we shall consider
both values TRUE and FALSE for the parameter KINCUT~\cite{charybdis2}.
\par
Overall, we expect the remnant BHs will have a typical speed $\beta_0=v_0/c$
with the distribution shown in the left panel of Fig.~\ref{beta}, for a sample of $10^4$ BHs,
where two different scenarios for the end-point of the decay were assumed.
The dashed line (KINCUT=TRUE) represents the case when the decay
is prevented from producing a remnant with proper mass $M_0$ below $M_{\rm G}$
(but could stop at $M_0>M_{\rm G}$), whereas the solid line (KINCUT=FALSE)
represents BH remnants produced when the last emission is only required to keep
$M_0>0$.
The mass $M_0$ for the remnants in these different cases is also distributed
according to the plots in the right panel of Fig.~\ref{beta}.
For KINCUT=TRUE, since the remnant mass $M_0\gtrsim M_{\rm G}$,
a smaller amount of energy is allowed to be emitted before the hole
becomes a remnant, whereas for KINCUT=FALSE much lighter remnants 
are allowed.
One could thus argue that the former scenario (KINCUT=TRUE) provides more
of a sensible description for BH remnants resulting from the partial decay of
quantum BHs than the latter (KINCUT=FALSE),
but we have employed both cases in our simulations for the sake of completeness.
\par
The same quantities, speed $\beta_0$ and mass $M_0$, but including just
the charged remnants, are displayed in Fig.~\ref{betac},
again for a sample of $10^4$ BH events. 
The left panel clearly shows that, including both scenarios,
one can expect the charged remnant velocity is quite evenly distributed
on the entire allowed range, but $\beta_0$ is generally smaller for KINCUT=TRUE.
\subsection{Remnant speed}
\begin{figure}[t]
\centerline{\begin{tabular}{cc}
\raisebox{3cm}{\small{$\beta_0$}}
\epsfig{figure=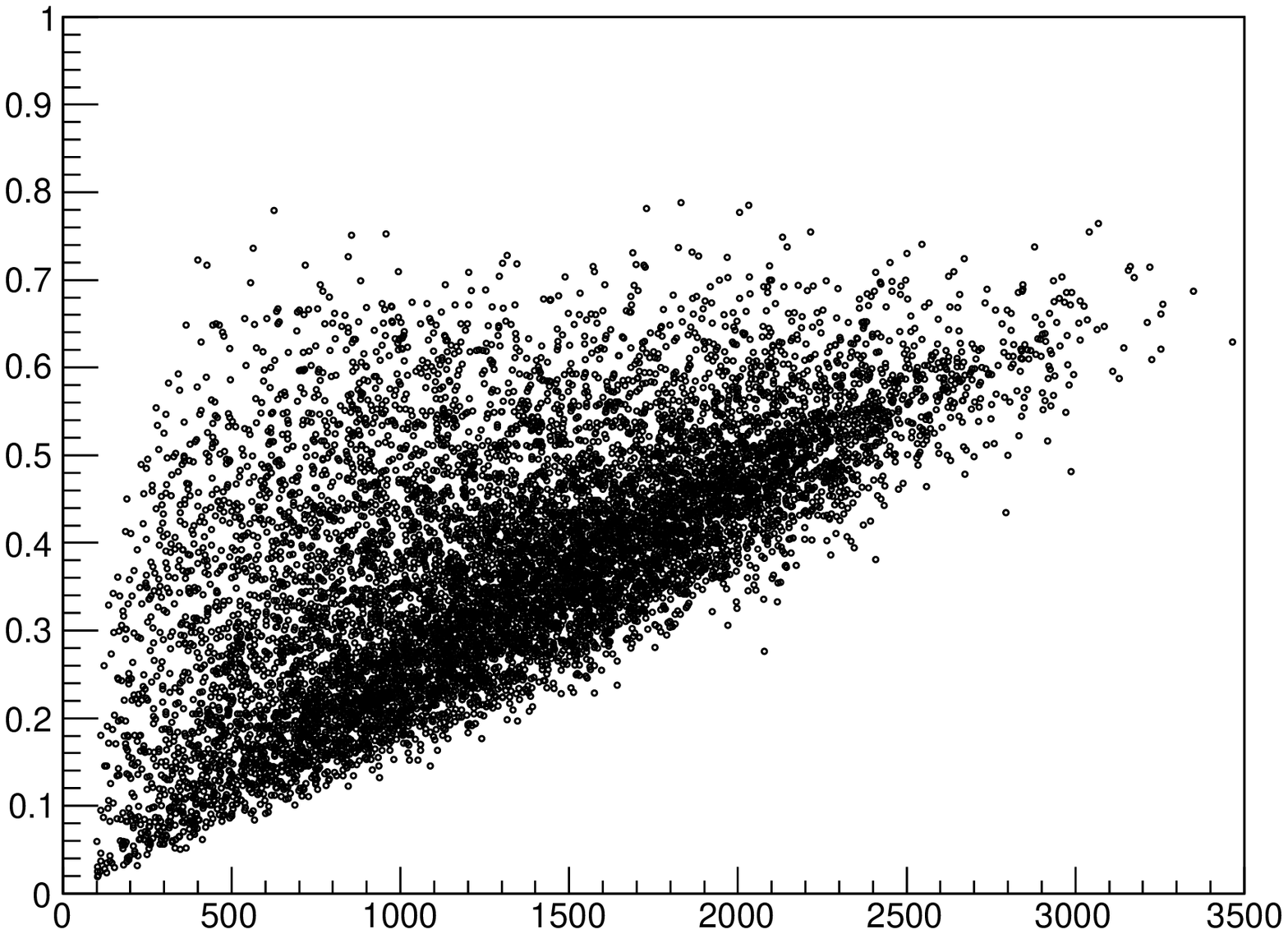,height=6cm,width=8cm}
&
\raisebox{3cm}{\small{$\beta_0$}}
\epsfig{figure=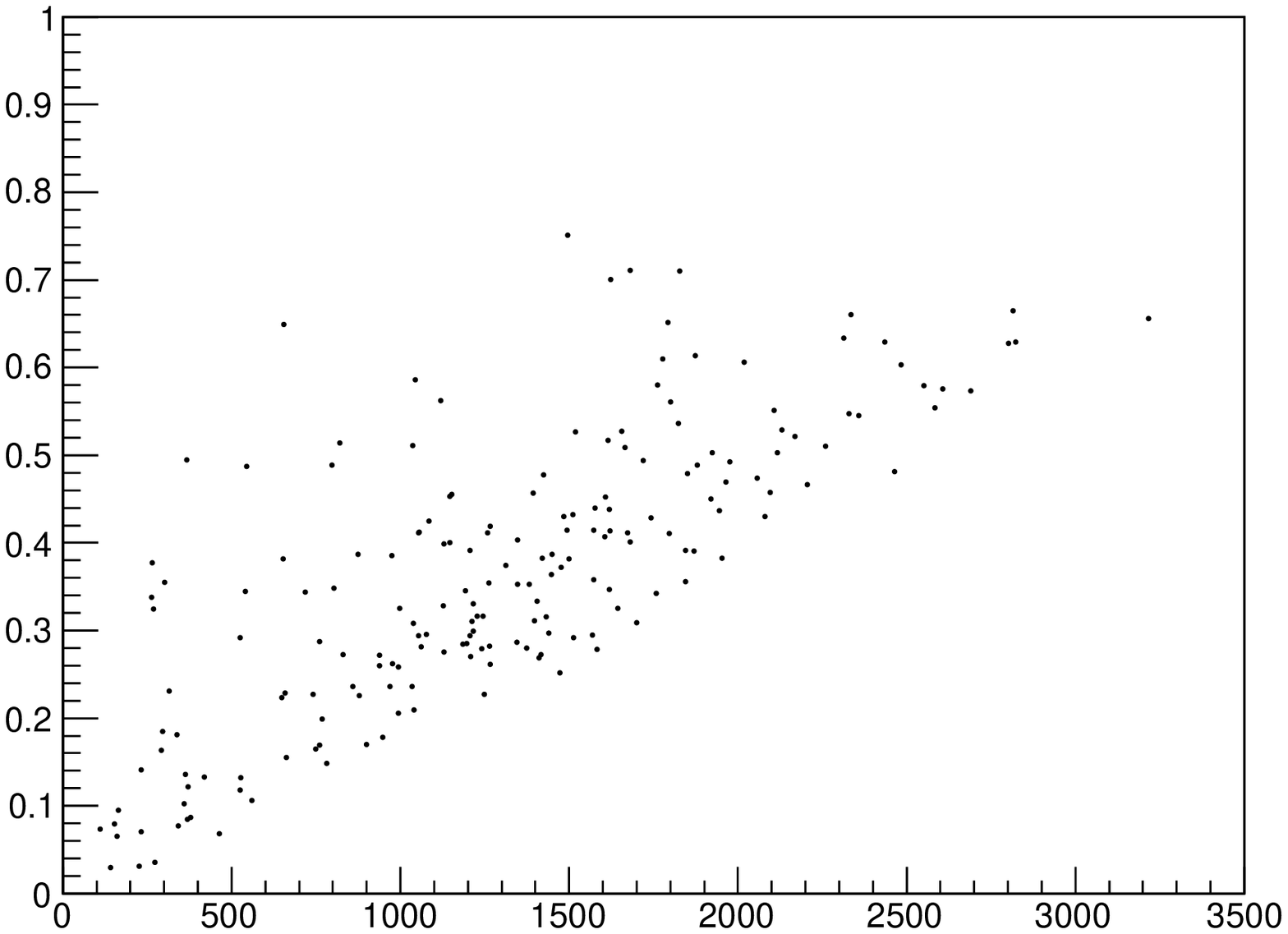,height=6cm,width=8cm}
\\
{\small{$P_T$}}
&
{\small{$P_T$}}
\end{tabular}}
\caption{Distribution of $\beta_0$ vs $P_T$ (in GeV) with KINCUT=TRUE for neutral remnants (left panel)
and charged remnants (right panel) for $P_T>100\,$GeV.
Both plots are for $\sqrt{s}=14\,$TeV with $M_{\rm G}=3.5\,$TeV in $D=6$ total dimensions
and $10^4$ total events.
\label{scatterBetaT}}
\end{figure}
\begin{figure}[t]
\centerline{\begin{tabular}{cc}
\raisebox{3cm}{\small{$\beta_0$}}
\epsfig{figure=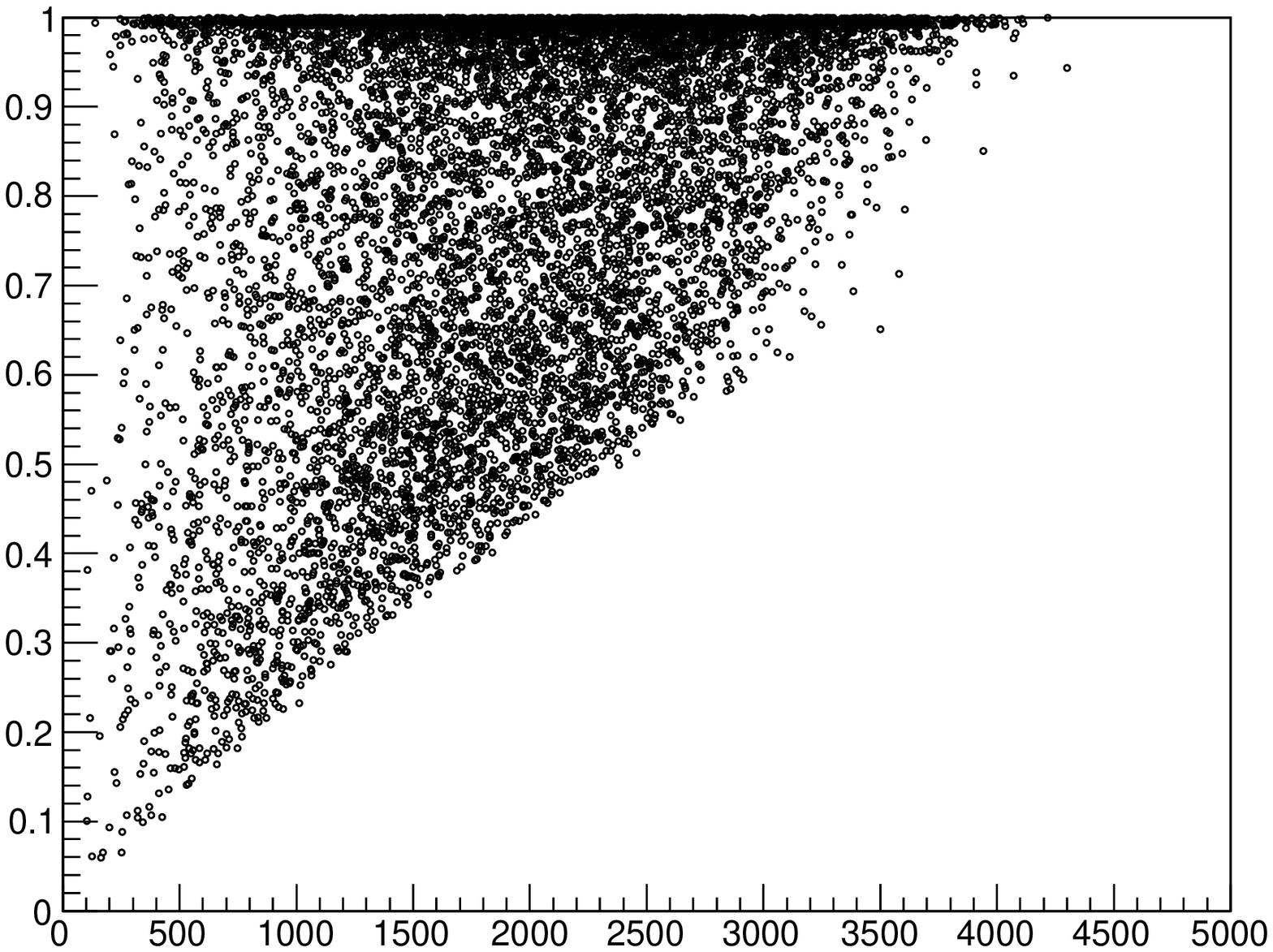,height=6cm,width=8cm}
&
\raisebox{3cm}{\small{$\beta_0$}}
\epsfig{figure=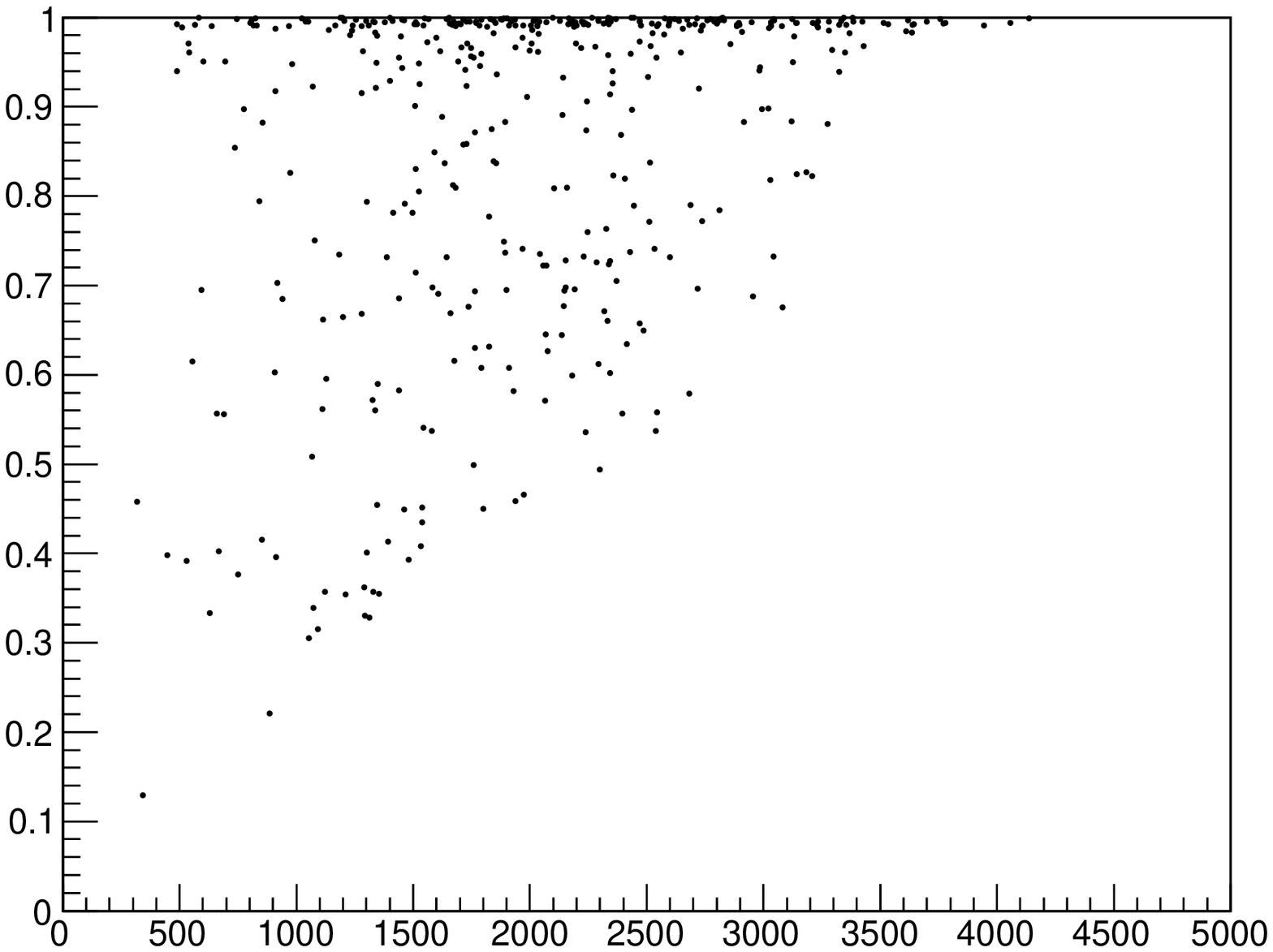,height=6cm,width=8cm}
\\
{\small{$P_T$}}
&
{\small{$P_T$}}
\end{tabular}}
\caption{Distribution of $\beta_0$ vs $P_T$  (in GeV) with KINCUT=FALSE for neutral remnants (left panel)
and charged remnants (right panel) for $P_T>100\,$GeV.
Both plots are for $\sqrt{s}=14\,$TeV with $M_{\rm G}=3.5\,$TeV in $D=6$
total dimensions and $10^4$ total events.
\label{scatterBetaF}}
\end{figure}
\begin{figure}[t!]
\centerline{\begin{tabular}{cc}
\raisebox{3cm}{\small{$\beta$}}
\epsfig{figure=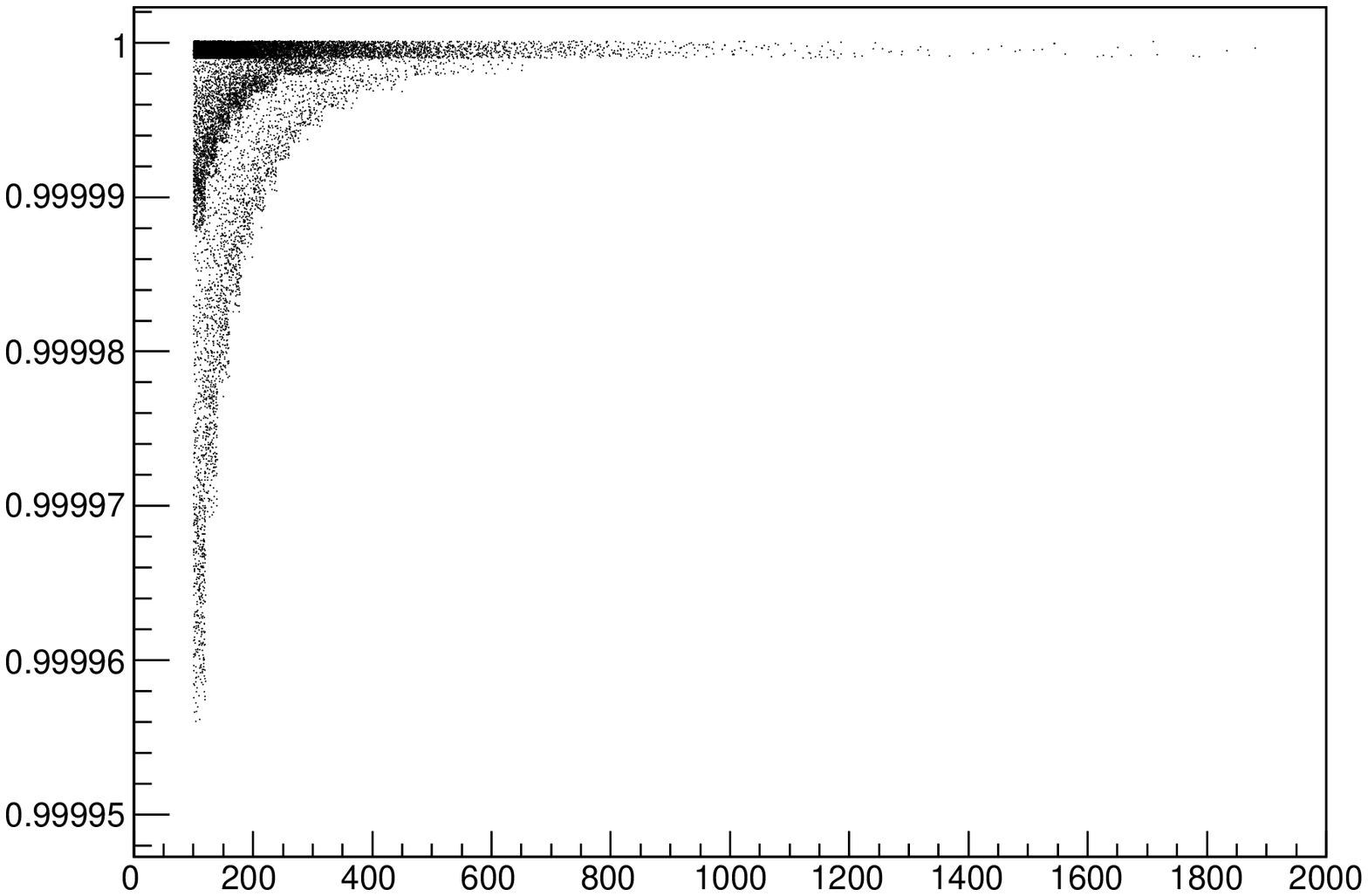,height=6cm,clip=}
&
\raisebox{3cm}{\small{$\beta$}}
\epsfig{figure=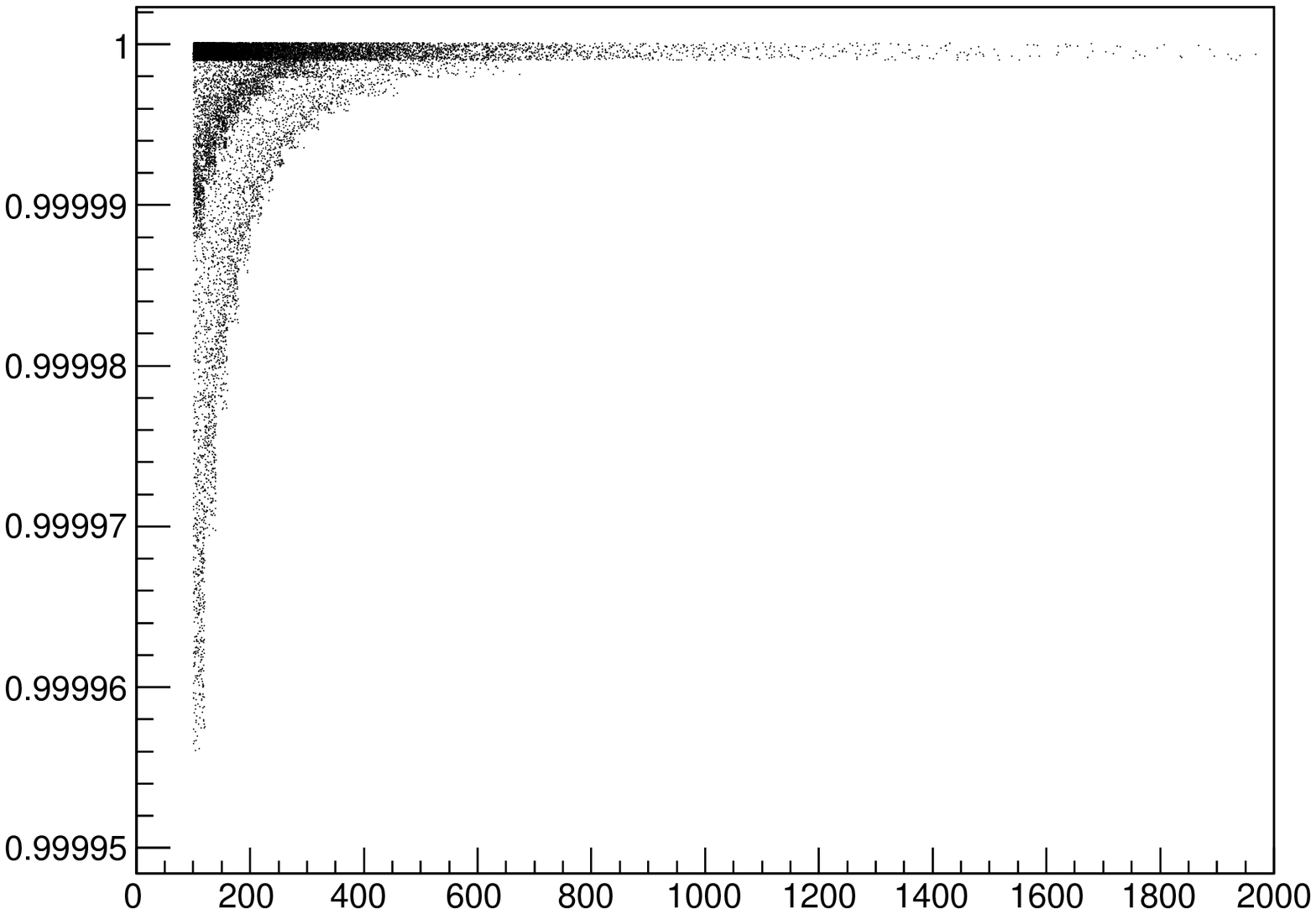,height=6cm,clip=}
\\
{\small{$P_T$}}
&
{\small{$P_T$}}
\end{tabular}}
\caption{Distribution of $\beta$ vs $P_T$ (in GeV) for background particles with $P_T>100\,$GeV,
in events with remnant BHs and KINCUT=TRUE (left panel) or KINCUT=FALSE (right panel).
Both plots are for $\sqrt{s}=14\,$TeV with $M_{\rm G}=3.5\,$TeV in $D=6$
total dimensions and $10^4$ total events.
\label{scatterBetaNo}}
\end{figure}

For phenomenological reasons, it is more instructive to consider the distribution of the speed
$\beta_0$ with respect to transverse momenta $P_T$ for remnant BHs with a cut-off $P_T>100\,$GeV,
and compare the same distribution for the ordinary particles produced in the same collisions
(the background particles).
Figs.~\ref{scatterBetaT} and~\ref{scatterBetaF} show, separately, the distributions
of $\beta_0$ for neutral and charged remnants, again for KINCUT=TRUE and FALSE,
respectively.
We first recall that the remnant velocities for KINCUT=TRUE are lower because the masses
of remnant BHs in this case are typically larger than the masses for KINCUT=FALSE.
Comparing then with the background particles in the same events, shown in Fig.~\ref{scatterBetaNo}, 
remnant BHs appear clearly distinguished for the more sensible choice of KINCUT=TRUE,
since there is hardly any BH with $\beta_0\gtrsim 0.7$, whereas all the background particles have
$\beta\simeq 1$.
For KINCUT=FALSE, the situation is somewhat less clear, since there are BHs with
large $\beta_0\gtrsim 0.9$, but a significant fraction of them still shows $\beta_0\lesssim 0.9$.
\par
\begin{figure}[h!]
\centerline{\begin{tabular}{c}
\raisebox{3cm}{\small{$\beta$}}
\epsfig{figure=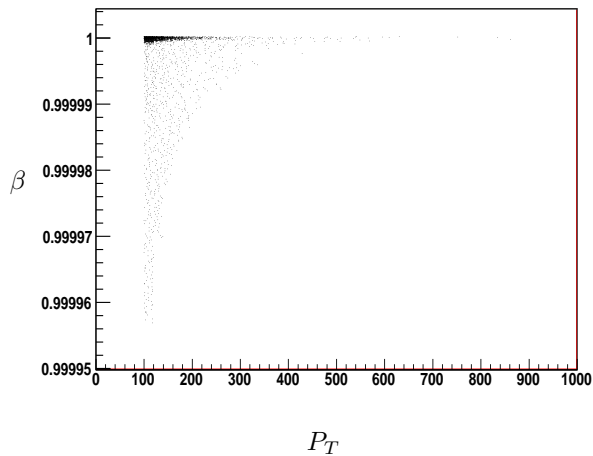,height=6cm,width=8cm}
\\
{\small{$P_T$}}
\end{tabular}}
\caption{Distribution of $\beta$ vs $P_T$ (in GeV) for particles with $P_T>100\,$GeV,
in events with $t\,\bar t$ for $\sqrt{s}=14\,$TeV.
\label{scatterBetatt}}
\end{figure}

Finally, the above speeds $\beta_0$ can be compared with the distributions of $\beta$ for the
$t\,\bar t$ process (which can be considered as one of the main backgrounds) shown in
Fig.~\ref{scatterBetatt}.
Let us also recall that the production cross section $\sigma_{t\,\bar t}(14\,{\rm TeV})\simeq 880\,$pb,
and the branching ratio for single-lepton decays we are here including as a background,
since we require final states with significant missing transverse energy, is $0.44$.
For a luminosity $L=10\,$fb$^{-1}$, we hence expect about $3.9\times 10^6$ such events.
This must be compared with the expected number of $400$ BH events that was mentioned
at the end of Section~\ref{production}, for the same luminosity.
\subsection{Energy release for charged remnants}
\label{bethe-bloch}
\begin{figure}[t!]
\centerline{\begin{tabular}{cc}
\raisebox{3cm}{\small{$\frac{dE}{dx}$}}
\epsfig{figure=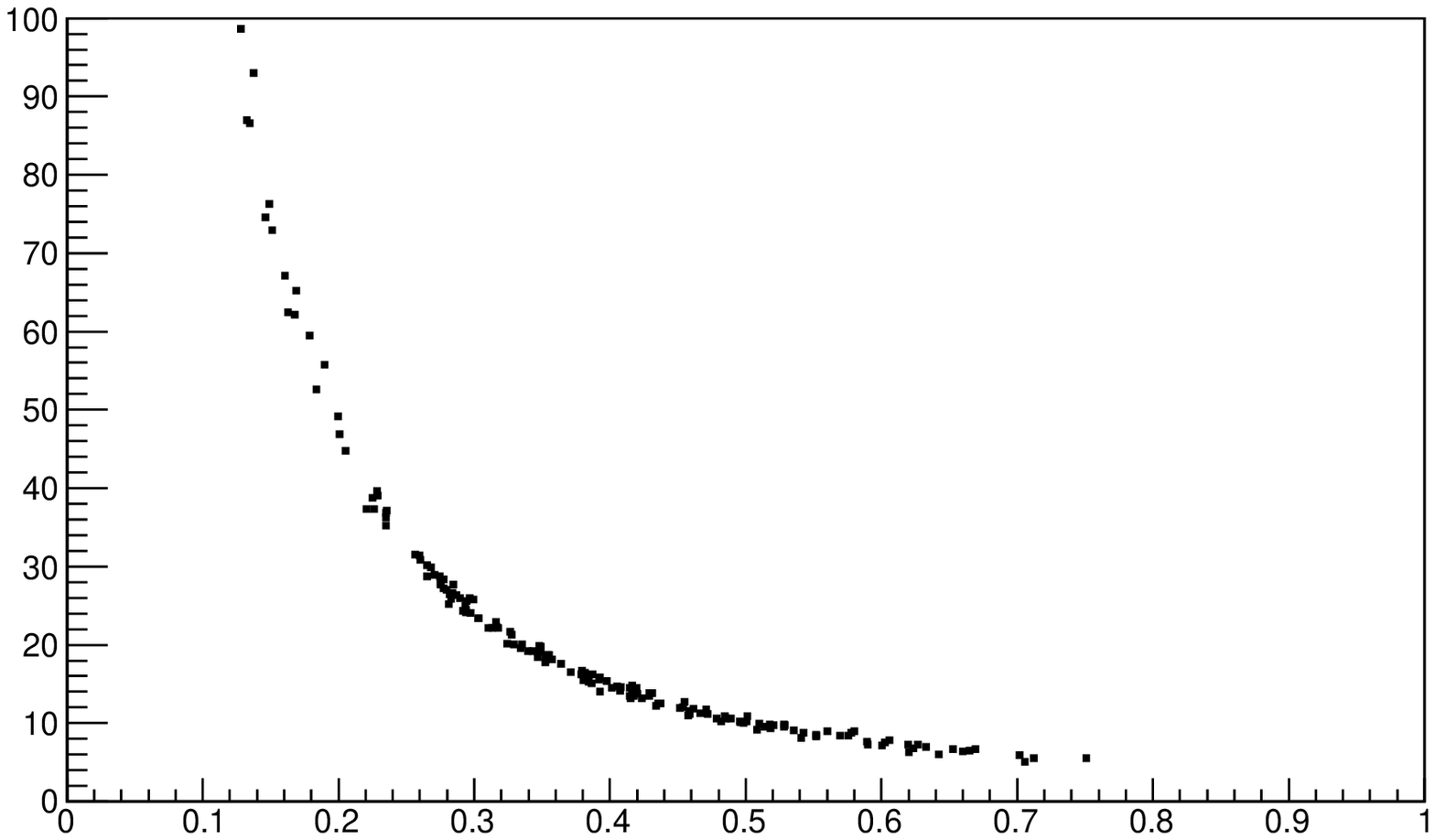,height=6cm,width=8cm}
&
\raisebox{3cm}{\small{$\frac{dE}{dx}$}}
\epsfig{figure=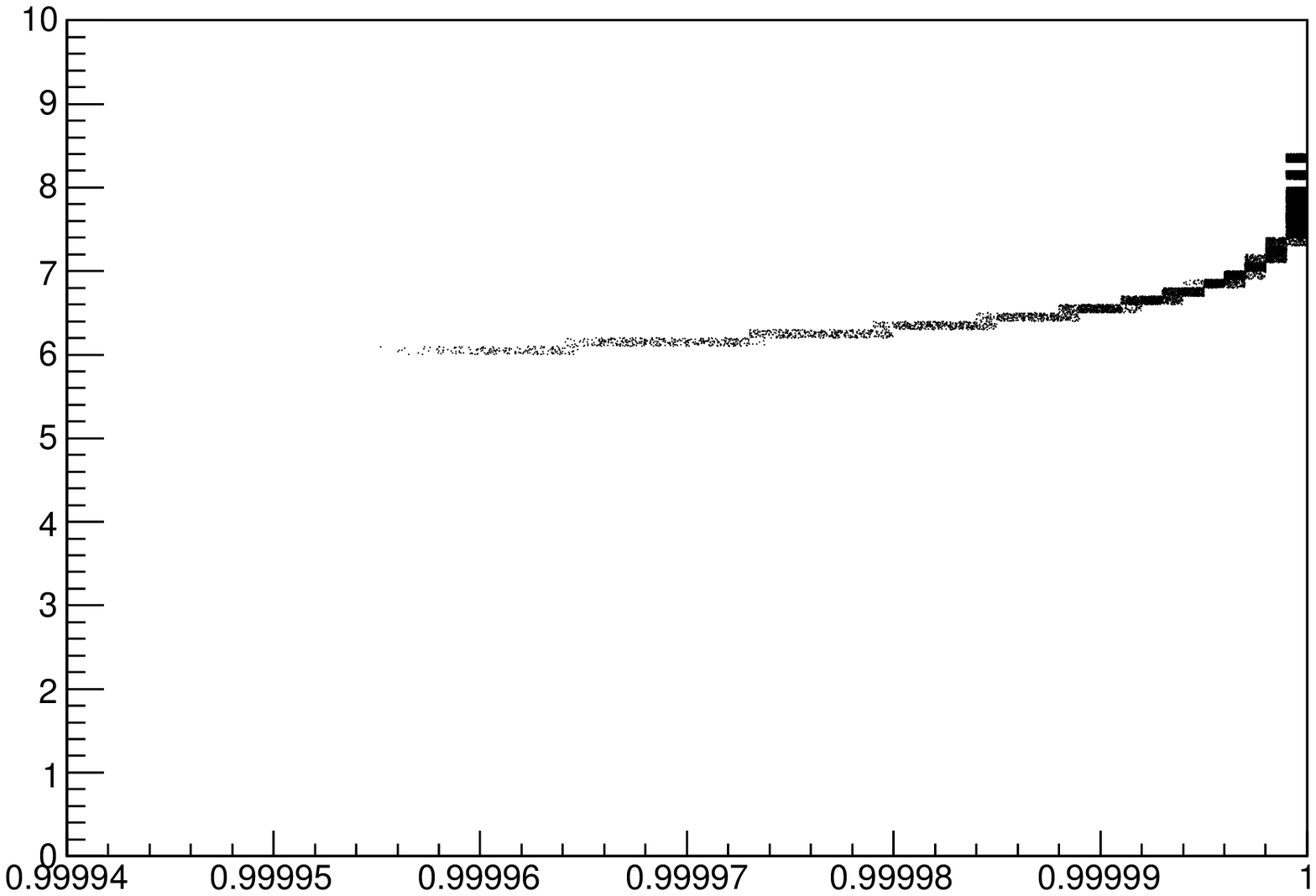,height=6cm,width=8cm}
\\
{\small{$\beta_0$}}
&
{\small{$\beta$}}
\end{tabular}}
\caption{Typical energy loss per unit distance (in MeV/cm) from charged remnant BHs 
vs $\beta_0$, for KINCUT=TRUE (left panel) and analogous quantity for background particles
(right panel).
Both plots are for $\sqrt{s}=14\,$TeV with $M_{\rm G}=3.5\,$TeV in $D=6$
total dimensions and $10^4$ total events.
}
\label{dEdxT}
\end{figure}
\begin{figure}[h!]
\centerline{\begin{tabular}{cc}
\raisebox{3cm}{\small{$\frac{dE}{dx}$}}\epsfig{figure=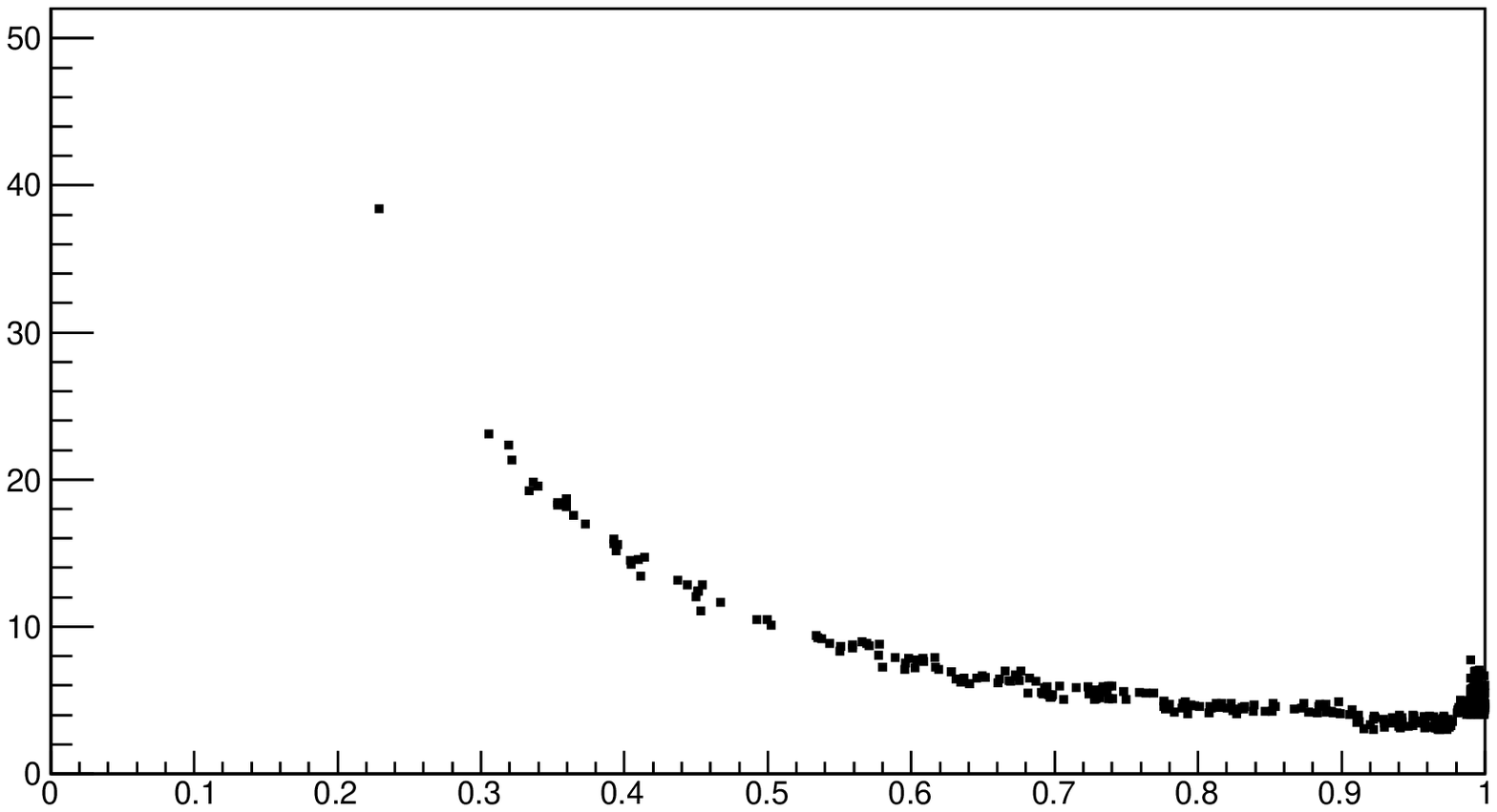,height=6cm,width=8cm}
&
\raisebox{3cm}{\small{$\frac{dE}{dx}$}}
\epsfig{figure=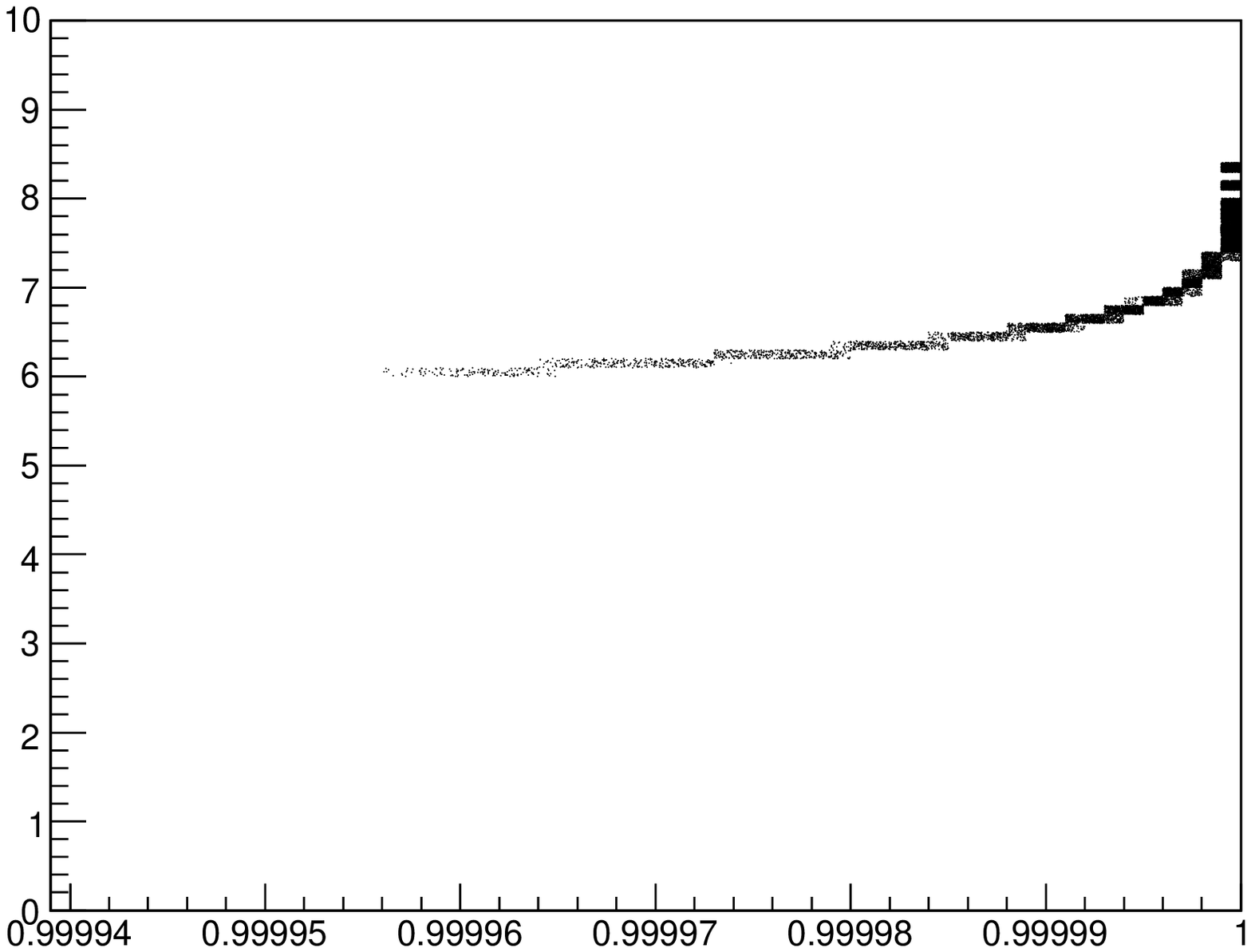,height=6cm,width=8cm}
\\
{\small{$\beta_0$}}
&
{\small{$\beta$}}
\end{tabular}}
\caption{Typical energy loss per unit distance (in MeV/cm) from charged remnant BHs 
vs $\beta_0$ for KINCUT=FALSE (left panel) and analogous quantity for background particles
(right panel).
Both plots are for $\sqrt{s}=14\,$TeV with $M_{\rm G}=3.5\,$TeV in $D=6$
total dimensions and $10^4$ total events.
}
\label{dEdxF}
\end{figure}
\begin{figure}[h!]
\centerline{\begin{tabular}{c}
\raisebox{3cm}{\small{$\frac{dE}{dx}$}}
\epsfig{figure=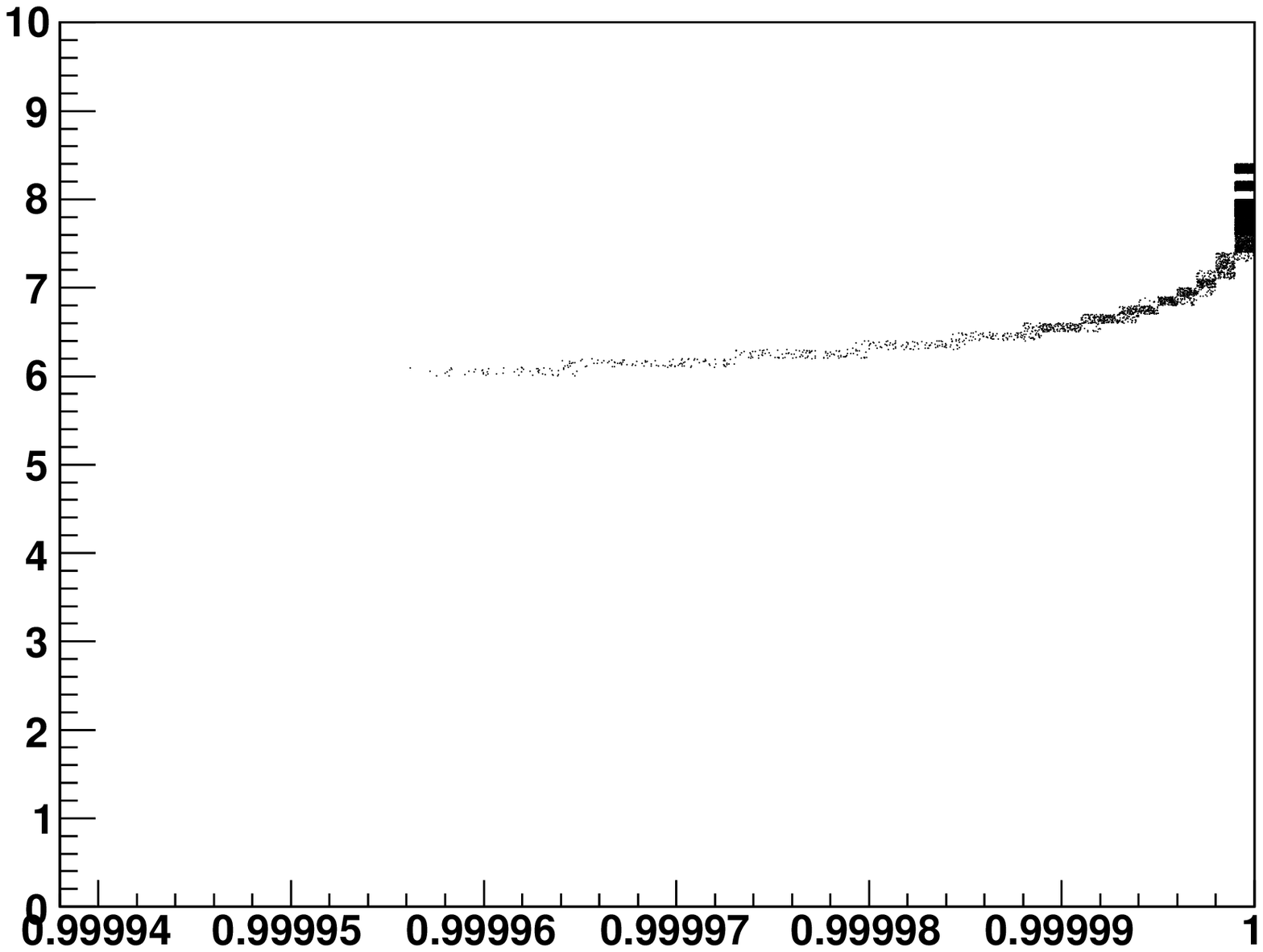,height=6cm,width=8cm}
\\
{\small{$\beta$}}
\end{tabular}}
\caption{Typical energy loss per unit distance (in MeV/cm) from charged particles vs $\beta$
in $10^4$ total events with $t\,\bar t$ at $\sqrt{s}=14\,$TeV.
}
\label{dEdxtt}
\end{figure}
The energy released in a medium by a particle of mass $M$ and charge
$Q=z\,e$ can be estimated using the well-known Bethe-Bloch equation.
In particular, for particle moving at relativistic speed, one has an energy loss
per distance travelled given by
\be
\frac{dE}{dx}
=
-4\,\pi\,N_A\,r_e^2\,m_e\,c^2\,
\frac{Z\,\rho}{A\,\beta^2}
\left[
\ln\left(\frac{2\,m_e\,c^2\,\beta^2}{I}\right)
-\beta^2-\frac{\delta}{2}\right]
\ ,
\label{bethe}
\ee
where $N_A$ is Avogadro's number, 
$m_e$ and $r_e$ the electron mass and classical radius,
$Z$, $A$ and $\rho$ the atomic number, atomic weight and density of the medium,
$I$ its mean excitation potential,
\be
I
\simeq
16\,Z^{0.9}\,{\rm eV}
\ ,
\ee
and $\delta$ a constant that describes the screening of the electric field
due to medium polarisation.
\par
For our case, we can use the values for Si, as the $dE/dX$ can be effectively
measured in the ATLAS Inner Detector, namely 
$\rho=2.33\,$g$/$cm$^3$, $Z= 14$, $A = 28$, $I = 172\,$eV and
$\delta=0.19$.
On using the $\beta_0$ for charged remnant BHs from the right panels of
Figs.~\ref{scatterBetaT} and~\ref{scatterBetaF}, one then obtains the typical
distributions displayed in Figs.~\ref{dEdxT} and Figs.~\ref{dEdxF}, where
the energy loss from remnant BHs is compared with analogous quantities
for ordinary particles coming from BH evaporation.
One can then also compare with the energy loss in $t\,\bar t$ events
displayed in Fig.~\ref{dEdxtt}.
From this comparison, we can see that for KINCUT=TRUE, a cut around
$10\,$MeV$/$cm would clearly isolate remnants BHs, since they would mostly
loose more energy.
Instead, for KINCUT=FALSE, the expected fraction of BHs loosing as much
is significantly smaller.
\par
Since we are only considering non-strongly interacting states, 
the charged stable remnants behave as massive muons, travelling long distances through
the detector and releasing only a negligible fraction of their total energy.
The main problem in detecting such states at the LHC is the trigger time width of $25\,$ns
(1 bunch crossing time).
Due to their low speed, most of them will reach the muon system out of time and could not
be accepted by the trigger.
A study performed at ATLAS in search for stable sleptons and R-hadrons~\cite{atlas_slep}
set a threshold cut of $\beta > 0.62$ in order to have a muon trigger in the event
(slower particles end up out of the trigger time window).
In order to access the low $\beta$ range, one can imagine to trigger on the missing transverse
energy ($E_T^{\rm miss}$), copiously produced by the charged remnants, or on other standard
particles produced in the BH evaporation (typically electrons or muons).
Regarding the missing transverse energy, from our simulations, we can infer that the 
efficiency of a cut at $100\,$GeV in calorimetric $E_T^{\rm miss}$~\footnote{The calorimetric
$E_T^{\rm miss}$ was evaluated considering $\nu$'s, gravitons, muons and charged remnants
as invisible particles.}
would be of the order of (more than) $90\%$ for the KINCUT=FALSE (TRUE) sample.
Another possibility is to trigger on ordinary particles, typically electrons or muons with
high transverse momentum $P_T$, in order to reduce the high potential background
coming from QCD multi-jet events.
A cut at $P_T>50\,$GeV would have an efficiency of about $50\%$,
both for the KINCUT=TRUE and KINCUT=FALSE samples.
Once the events have been accepted by the trigger the signal has to be isolated from
the background by means of the $dE/dX$ measurement.
Technical issues due to a saturation effect in such a measurement will limit the 
range of accessible $\beta$ towards low values at increasing $dE/dX$.
For example the ATLAS experiment reported a limit $\beta \gtrsim 0.2$~\cite{atlas_slep}.
\section{Conclusions and outlook}
\label{conc}
\setcounter{equation}{0}
We have investigated the possibility that events with the production of remnant BHs
leave a distinct signature at the LHC.
In particular, we have simulated the generation of such objects using the Monte Carlo
code CHARYBDIS2 and analysed the distribution of the possible remnants' speed $\beta_0$.
We found that BHs could be produced with $\beta_0$ much lower than typical $\beta\simeq 1$
of SM particles.
Moreover, for the charged remnants, we also estimated the energy release inside the
detector, which could turn out to be significantly larger than the one expected from
SM particles.
\par
The main LHC experiments are designed to detect SM charged particles
which are produced with velocities $\beta=v/c$ large enough to fall into the LHC trigger window
of $25\,$ns.
For particles at the LHC to be detected and associated with the correct bunch crossing,
they have to be seen at most $25\,$ns after the arrival time of the particles which travel
at the speed of light~\cite{Kraan:2005ji,Hauser:2004nd}.
Later arrival would imply triggering or detection within future crossing time windows,
which implies a minimum relativistic factor $\beta_{\rm min}\sim 0.6$ for the ATLAS detector
and a bit less for the CMS detector, which is more compact in size.
Combined with the simulations, which showed the even distribution of the remnant velocity
on the entire allowed range, this means that there is a significant fraction of BHs
which cannot be triggered by the current LHC experiments.
As discussed in section~\ref{bethe-bloch}, it would still be possible to access lower values
of $\beta$ by triggering on the missing transverse energy or on ordinary particles
(typically high transverse momentum electrons or muons) produced in association
with the remnant in the BH evaporation.
\par
We would like to conclude by mentioning that the MoEDAL
(Monopole and Exotics Detector at the LHC) experiment~\cite{Pinfold:2010zza}
will complement the existing experiments.
This detector should be able to track electrically charged stable massive particles
with $Z/\beta$ as low as five (where $Z$ is the electric charge of the particle).
The MoEDAL experiment will be using the passive plastic track technique,
which does not require a trigger, and represents an excellent method to detect
and accurately measure the tracks of highly ionising particles and also their $Z/\beta$. 
\section*{Acknowledgements}
G.L.~A.~would like to thank L.~Fabbri and R.~Spighi for the very useful
discussions.
This work is supported in part by the European Cooperation
in Science and Technology (COST) action MP0905 ``Black Holes in a Violent  Universe".
The work of X.~C.~is supported in part by the Science and Technology Facilities Council
(grant number~ST/J000477/1).
O.~M.~is supported by UEFISCDI grant PN-II-RU-TE-2011-3-0184. 
\appendix
\section{Charged remnant BHs in the Brane-World}
\label{BWBH}
\setcounter{equation}{0}
A simple application of the four-dimensional Reissner-Nordstr\"om
metric to BHs with mass $M\sim M_{\rm G}$ and charge $Q\sim e$ would show
that such objects must be naked singularities.
However, in brane-world models, one can employ the tidal charge form of the
metric and find that, provided the tidal charge $q$ is strong enough, microscopic
BH can carry a charge of the order of $e$~\cite{CH}.
In particular, the horizon radius is now given by
\be
R_{\rm H}
=
\ell_{\rm P}\,
\frac{M}{M_{\rm P}}
\left(
1+
\sqrt{1-\tilde Q^2\,\frac{M_{\rm P}^2}{M^2}
+\frac{q\,M_{\rm P}^2}{\ell_{\rm G}^2\,M^2}}
\right)
\ ,
\label{tidalH}
\ee
where $M_{\rm P}$ and $\ell_{\rm P}$ are the Planck mass and length, respectively,
and $\tilde Q$ is the electric charge in dimensionless units, that is
\be
\tilde Q
\simeq
10^8\left(\frac{M}{M_{\rm P}}\right)
\frac{Q}{e}
\ .
\ee
Reality of Eq.~\eqref{tidalH} for a remnant of charge $Q=\pm e$
and mass $M\simeq M_{\rm G}$ then requires
\be
q
\gtrsim
10^{16}\,\ell_{\rm G}^2
\left(\frac{M_{\rm G}}{M_{\rm P}}\right)^2
\sim
10^{-16}\,\ell_{\rm G}^2
\ .
\ee
Configurations satisfying the above bound were recently found in 
Ref.~\cite{covalle}.
\end{document}